\documentclass[12pt]{article}
\usepackage{graphicx}
\usepackage{amsmath}
\usepackage{amssymb}

\begin{document}

\title{Dark Energy and Electrons}
\date{}
\author{B.G.Sidharth\footnote{sidharth@fias.uni-frankfurt.de} \footnote{Permanent address:B.M. Birla Science Center Adarsh Nagar, Hyderabad India}\\ Frankfurt Institute for Advanced Studies\\
Johann Wolfgang Goethe University\\ Ruth-Moufang-Str. 1\\ 60438 Frankfurt am Main Germany}
\maketitle

\begin{abstract}
In the light of recent developments in Dark Energy, we consider the electron in a such a background field and show that at the Compton wavelength the electron is stable, in that the Cassini inward pressure exactly counterbalances the outward Coulomb repulsive pressure thus answering a problem of the earlier electron theory.
\end{abstract}

\section{Introduction}\label{sec-1}
For more than a decade now, Dark Energy has been the ruling paradigm. It is now agreed that this mysterious Dark Energy is ubiquitous and seems to provide a small acceleration to the universe as a whole. It maybe mentioned that the author's 1997 model \cite{bg-cu01,bgmg8-99,bg-ijmpa98,bg-ijtp98} correctly predicted all this which was then confirmed by the observations of distant supernovae by Perlmutter and others in 1998. The background Dark Energy was identified with the Quantum vacuum with its Zero Point field \cite{duffy}. In several papers it was further pointed out that elementary particles like electrons would ``condense'' out of this background in a phase transition sense, rather like the formation of B\'{e}nard cells \cite{nipr-89}. \\ We will now examine all this in some detail particularly the important question of the stability of the particles -- the electrons -- so formed in the background Dark Energy.
We may reiterate that the ``mysterious'' background
Dark Energy is the same as the Quantum Zero Point fluctuations in
the background vacuum electromagnetic field .The background Zero
Point Field is a collection of ground state oscillators \cite{mwt}.
The probability amplitude as is known is
$$\psi (x) = \left(\frac{m\omega}{\pi \hbar}\right)^{1/4} e^{-(m\omega/2\hbar)x^2}$$
for displacement by the distance $x$ from its position of
equilibrium. So the oscillator fluctuates over an interval
$$\Delta x \sim (\hbar/m\omega)^{1/2}$$
The background \index{electromagnetic}electromagnetic field is an
infinite collection of independent oscillators, with amplitudes
$X_1,X_2$ etc. The probability for the various oscillators to have
amplitudes $X_1, X_2$ and so on is the product of individual
oscillator amplitudes:
$$\psi (X_1,X_2,\cdots ) = e^{ [-(X^2_1 + X^2_2 + \cdots)]}$$
wherein there would be a suitable normalization factor. This
expression gives the probability amplitude $\psi$ for a
configuration $B (x,y,z)$ of the magnetic field that is described by
the Fourier coefficients $X_1,X_2,\cdots$ or directly in terms of
the magnetic field configuration itself by, as is known,
$$\psi (B(x,y,z)) = P e^{ \left(-\int \int \frac{\bf{B}(x_1)\cdot \bf{B}(x_2)}{16\pi^3\hbar cr^2_{12}} d^3x_1 d^3x_2\right)}.$$
$P$ being a normalization factor. At this stage, it must be emphasized, we are thinking in
terms of energy without differentiation, that is, without
considering Electromagnetism or Gravitation as being separate as will be seen below. Let us
consider a configuration where the field is everywhere zero except
in a region of dimension $l$, where it is of the order of $\sim
\Delta B$. The probability amplitude for this configuration would be
proportional to
$$e^{ [-((\Delta B)^2 l^4/\hbar c)]}$$
So the energy of \index{fluctuation}fluctuation in a region of
length $l$ is given by finally, the density \cite{mwt,uof}
$$B^2 \sim \frac{\hbar c}{l^4}$$
So the energy content in a region of volume $l^3$ is given by
\begin{equation}
Energy\sim \hbar c/l\label{4e1}
\end{equation}At this stage, we note that if in (\ref{4e1}), we take $l$ to be the Planck length, then the energy turns out to be
$m_P c^2,$ where $m_P$ is the Planck mass. This energy can be considered to be entirely gravitational because as is well known  at the Planck scale, $Gm_P^2 \sim e^2.$
It has been shown that this constitutes the limit as at higher energies, the oscillations become totally chaotic \cite{ce-98,cegasc-72}.\\
As an alternative derivation, it is interesting to derive a model
based on the theory of phonons which are quanta of sound waves in a
macroscopic body \cite{huang}. Phonons are a mathematical analogue
of the quanta of the electromagnetic field, which are the photons,
that emerge when this field is expressed as a collection of Harmonic
oscillators. This situation is carried over to the theory of solids
which are made up of atoms that are arranged in a crystal lattice
and can be approximated by a sum of Harmonic oscillators
representing the normal modes of lattice oscillations. In this
theory, as is well known the phonons have a maximum frequency
$\omega_m$ which is given by
\begin{equation}
\omega_m = c \left(\frac{6\pi^2}{v}\right)^{1/3}\label{4e2}
\end{equation}
In (\ref{4e2}) $c$ represents the velocity of sound in the specific
case of phonons, while $v = V/N$, where $V$ denotes the volume and
$N$ the number of atoms. In this model we write
$$l \equiv \left(\frac{4}{3} \pi v\right)^{1/3}$$
$l$ being the inter particle distance. Thus (\ref{4e2}) now becomes
\begin{equation}
\Omega \equiv\omega_mx=c/l\label{4e3}
\end{equation}
Let us now liberate the above analysis from the immediate scenario
of atoms at lattice points and quantized sound waves due to the
Harmonic oscillations and look upon it as a general set of Harmonic
oscillators as above. Then we can see that (\ref{4e3}) and
(\ref{4e1}) are identical as
\begin{equation}\label{4e4}
 \omega = \frac{mc^2}{\hbar}
\end{equation}
\section{The Electron}
Let us now consider the electron. Historically the original concept of the
electron was that of a spherical charge distribution
\cite{rohr,barut,jc}. It is interesting to note that in the
non-relativistic case, it was originally shown that the entire
inertial mass of the electron equalled its electromagnetic mass.
This motivated much work and thought in this interesting direction.
To put it briefly, in non relativistic theory, we get \cite{rohr},
$$\mbox{Kinetic \, energy} =  (\beta /2) \frac{e^2}{Rc^2} v^2,$$
where $R$ is the radius of the electron and $\beta$ is a numerical
factor of the order of $1$. So we could possibly speak of the entire
mass of the electron in terms of its
electromagnetic properties that is without invoking new parameters like mass separately into the theory. \\
It might be mentioned that it was still possible to think of an
electron as a charge distribution over a spherical shell within the
relativistic context, as long as the electron was at rest or was
moving with a uniform velocity. However it was now necessary to
introduce, in addition to the electromagnetic force, the Poincare
stresses - these were required to
counter balance the mutual Coulombic repulsive ``explosion'' of the different parts of the electron the so called self energy.
When the electron in a field is accelerated, the above picture no
longer holds. We have to introduce the concept of the electron self
force which as is known, is given by, in the simple case of one
dimensional motion,\cite{bgs-tlsr->}
\begin{equation}
F = \frac{2}{3} \frac{e^2}{c^3} \frac{e^2}{Rc^2} \ddot{x} -
\frac{2}{3}\frac{e^2}{c^3} \frac{d}{dt} \ddot{x} + \gamma \frac{e^2
R}{c^4} \ddot{x} + \beta 0 (R^2)\label{ex}
\end{equation}
where dots denote derivatives with respect to time, and $R$ as
before is the radius of the spherical electron and $\beta$ is merely
a dimensional factor and $\gamma$ is a numerical coefficient of the
order one. Interestingly the coefficient of the second term on the
right hand side is independent, not only of the charge distribution,
but also of the assumed radius $R$.\\ More generally (\ref{ex})
becomes a vector equation. In (\ref{ex}), the first term on the
right side gives the electromagnetic mass of the earlier theory. As
can be seen from (\ref{ex}), as $R$ the size of the electron $\to 0$
the first term $\to \infty$ and this is a major inconsistency. In
contrast the second term which contains the non Newtonian third time
derivative remains unaffected while the third and following terms
$\to 0$. It may be mentioned that the first term (which $\to
\infty$) gives the electromagnetic mass of the electron while the
second term gives the well known Schott term
(Cf.ref.\cite{rohr,barut,bgs}). Its presence is required however
because it compensates the energy loss due to radiation by the
accelerated electron.\\ In any case it is possible to develop a model
of an extended electron consistent with relativity on these lines,
but at the expense of introducing non electromagnetic
forces. However it has been shown over the years that with the introduction of minimum space time intervals, typically at the Compton scale all these problems are overcome \cite{diracpqm,bgs-tu08}. The minimum space time interval removes, firstly some unphysical effects like the advanced non-causal field of the earlier theory
because these take place within the Compton time and secondly the
infinite self energy of the point electron disappears due to the
Compton scale. \\The Compton scale comes as a Quantum Mechanical effect, within which
we have Zitterbewegung effects and a breakdown of Causal Physics
\cite{diracpqm}. Indeed Dirac had noted this aspect in connection
with two difficulties with his electron equation. Firstly the speed
of the relativistic Quantum Mechanical electron turns out to be the
velocity of light. Secondly the position coordinates become complex
or non Hermitian. His explanation was that in Quantum Theory we
cannot go down to arbitrarily small space time intervals, for the
Heisenberg Uncertainty Principle would then imply arbitrarily large
momenta and energies. So Quantum Mechanical measurements are an
average over intervals of the order of the Compton scale. Once this
is done, we recover meaningful physics. All this has been studied
afresh by the author more recently, in the context of a non
differentiable space time and noncommutative geometry \cite{uof,bgs-tu08}.
In Classical Physics the point electron leads to infinite self energy via the
electromagnetic mass term $ \propto e^2/R$, where $R$ is the radius which is
made to tend to zero. If on the other hand $R$ does not vanish, in
other words we have an extended electron, then we have to introduce
non electromagnetic forces like the Poincar\'{e} stresses for the
stability of this extended object, though on the positive side this
allows the radiation damping or self force that is required by
conservation laws \cite{bgs-tlsr->}.\\
Dirac could get rid of these problems by introducing the difference
between the advanced and retarded potentials in his phenemenological
equation in which the infinity was absorbed into a renormalized
point particle mass: This was the content of the classical Lorentz-Dirac
equation. A new term represents the radiation damping effect, but
we then have to contend with the advanced potential or equivalently
a non locality in time \cite{jica-99,hoyle}. However this non locality takes place within
the Compton time within which the
electron attains a luminal velocity.\\
So the Lorentz-Dirac equation on the other hand had unsatisfactory
features like the derivative of the acceleration, the non locality
in time and the run away solutions, features confined to the Compton scale.\\
Feynman and Wheeler bypassed the infinity and the extended
electron self force - but the mass was no longer electromagnetic.
Moreover the net result is that there is only the desired retarded
potential \cite{bgs-tlsr->}. But an instantaneous interaction with the rest of the
charges of the universe is required. It is this interaction with the
remaining charges which leads to the point electron's self energy.
Surprisingly however the interaction with the rest of the charges in
the immediate vicinity of the given charge in the Feynman-Wheeler
formulation gives us back the Dirac term in the Lorentz-Dirac equation with
its non locality within the Compton scale. There is thus a
reconciliation of the Dirac and the Feynman Wheeler approaches, once
we bring into the picture, the Compton scale.\\
Outside this scale, the theory is causal that is uses only the physical
retarded potential because effectively the  unphysical advanced potential gets
canceled out as it appears as the sum of the symmetric and
antisymmetric differences.\\
The final conclusion was that in a Classical context a totally
electromagnetic electron is impossible as also the concept of a
point electron without introducing additional ``unphysical'' concepts
including action at a distance. It was believed therefore that the
electron was strictly speaking the subject of Quantum Theory.\\
Nevertheless in Dirac's relativistic Quantum Electron, we again
encounter the electron with the luminal velocity within the Compton
scale, precisely what was encountered in Classical Theory as well,
as noted above. This again is the feature of a point space time
approach. At this stage a new input was given by Dirac as noted - meaningful
physics required averages over the Compton scale, in which process,
the unphysical zitterbewegung effects were eliminated. Nor has
Quantum Field Theory solved the problem - one has to take recourse
to renormalization, and as pointed out by Rohrlich, one still has a
non electromagnetic electron. In any case, it appears that further
progress would come either from giving up point space time or from
an electron that is extended (or has a sub structure) in some sense
\cite{rohr,barut,jc,hoyle}. \\ From this point of view the relativistic
theory of the electron is inconclusive to date. As noted by Feynman
himself in his famous Lectures on Physics (Vol II), ``We do not know
how to make a consistent theory - including the Quantum Mechanics -
which does not produce an infinity for the self energy of the
electron, or any point charge. And at the same time there is no
satisfactory theory that describes a non-point charge...'' In the
words of Hoyle and Narlikar \cite{nar}, ``...it was believed that the
problem of the self force of the charge would not be solved except
by recourse to Quantum Theory... This hope has not been fully
realized. Quantum Field Theory does alleviate the self energy
problem but cannot surmount it without introducing the
renormalization programme...'' In renormalization of Quantum Field Theory this is overcome by adding an ad hoc  infinite negative mass term to compensate the infinite positive Coulomb term. All this appears sufficiently ad hoc as to merit a search for an alternative.Infact Dirac himself expressed the opinion that the renormalization programme was a fluke and that it would be discarded one day \cite{uof}. \section{The Electron's Stability}
With the above background let us consider the stability of the electron that is formed by such a ``condensation''  from a background Quantum vacuum as in (\ref{4e1}). We consider the shell model of the electron that occupied the center stage for the early decades of the twentieth century \cite{bgs-fw->,bgs-tlsr->}.
This model was much later re-examined  by Casimir in the light of the possibility that the Casimir energy may provide Poincar\'{e} stresses required in the early model to retain the stability of the electron \cite{motr-88,puthoff}.To sum up Casimir had suggested that the background Zero Point field can manifest itself in the
force of attraction for example between two thin metallic plates in vacuum as was experimentally verified.\\ Based on these ideas
Casimir suggested that a dense shell-like distribution of charge might suppress vacuum fields in the interior. This could result in inward radiation
pressure from the electromagnetic vacuum fluctuation fields thus compensating outwardly-directed Coulomb forces to yield a stable
configuration at small dimensions.We will examine in a little detail how the model of an electron as an extended object
at the Compton scale as above is self consistent.
 Let us consider the charge $e$ to be homogeneously distributed on a spherical shell of
radius $ R.$
 So the electric field $E$ is given by
\begin{equation}\label{puthoff1}
    E = \frac{e}{4 \pi \varepsilon_0 r^2},
\end{equation}
This leads to the  Coulomb energy \cite{puthoff}

\begin{equation}\label{puthoff2}
    E_{coul}  = \int_R^\infty \frac{1}{2}
    \varepsilon_0 E^2 dV =  \left(\frac{\alpha \hbar c}{2 R}\right),
\end{equation}
This is the infinite self energy of the point electron in the limit as $R \rightarrow 0.$
In (\ref{puthoff2}) $\alpha $ is the fine structure constant, $\alpha = e^2/4\pi \varepsilon_0 \hbar c \approx 1/137$
 With regard to the vacuum fluctuation electromagnetic fields, the spectral energy
density is given by the well known expression
 \[\rho(\omega) d\omega = \frac{\hbar \omega^3}{2\pi^2 c^3}d\omega\]

This leads to the energy density
\begin{equation}\label{puthoff4}
    u_{vac} = \left ( \int_0^{\Omega} \frac{\hbar \omega^3}{2 \pi^2c^3}d\omega \right )= \left (\frac{\hbar \Omega^4} {8\pi^2c^3}\right )
\end{equation}
In (\ref{puthoff4}) $\Omega \equiv  \omega_{max}$ constitutes an upper limit cutoff frequency. The vacuum energy inside the sphere is now given by $R$
\begin{equation}\label{puthoff5}
    E_{vac}=-u_{vac} V= \left ( -\frac{\hbar \Omega ^4}{8 \pi^2c^3}.\frac{4}{3}\pi R^3 \right )
    =  \left ( -\frac{\hbar \Omega ^4 R^3}{6 \pi c^3} \right ).
\end{equation}
From (\ref{puthoff2}) and (\ref{puthoff5}) we get,
\begin{equation}\label{puthoff6}
    E=E_{coul}+E_{vac}=\left ( \frac{\alpha \hbar c}{2R} \right ) -
     \left ( \frac{\hbar \Omega ^4 R^3}{6 \pi c^3} \right )
\end{equation}
The outwardly-directed Coulomb pressure is given by

\begin{equation}\label{puthoff7}
    P_{coul}=u_{coul}= \left ( \frac{\alpha \hbar c}{8 \pi R^4}\right ),
\end{equation}
while the inwardly-directed vacuum radiation pressure as is known is \cite{dewi}
 \begin{equation}\label{puthoff8}
    P_{vac}= -\frac{1}{3}u_{vac}= \left ( -\frac{\hbar \Omega ^4}{24 \pi^2c^3} \right ).
 \end{equation}

  We require for stability that (\ref{puthoff7}) and (\ref{puthoff8}) cancel each other.This happens for $R= l$ given by
  \begin{equation}\label{puthoff9}
   l =  \left [ \frac{(3\pi \alpha)^{1/4}c}{\Omega}\right ]
  \end{equation}
  The energy $E$ in (\ref{puthoff6}) then reduces to
  \begin{equation}\label{puthoff10}
     E= E_{coul}+E_{vac}=\left ( \frac{\alpha \hbar c}{2l} \right )-
      \left ( \frac{\alpha \hbar c}{2l} \right )\equiv 0,
  \end{equation}
It can be seen that (\ref{puthoff9}) is the same as (\ref{4e3}) because $\Omega \equiv \omega_{max}$
In other words invoking (\ref{4e4}), it can be seen from (\ref{puthoff9}) that $l$ is the Compton wave length $l=\frac{\hbar}{2mc}.$\\
At the Compton wave length the Coulomb self energy (which diverged in the limit $R \rightarrow 0$ in the old theory) is balanced  by the background vacuum energy, and we have a stable configuration.This justifies the  approach in Section \ref{sec-1}. In other words, we can consider the electron to be formed by the mechanism leading to (\ref{4e1}), with $l$ being the electron's Compton wavelength. Such an electron has been shown to be a stable configuration.\section{Remarks}
We finally make a few comments. Einstein himself believed that the electron was some sort of a condensate from a background electromagnetic field. Indeed he even observed that it was sufficient to understand the electron.\\ It is interesting to note that the fluctuation of the metric is given by \cite{mwt} $$ \Delta g \sim l_P/l^3,$$ where $l_P$ is the Planck length. This is for practical purposes extremely small. However when $l$ equals the Compton wavelength, as in our case,then $\Delta g $ is of the order of 1, that is it is considerable, pointing to a significant departure in behavior. Indeed, in the completely different context of quantized strings, Veneziano too noted \cite{veneziano} that  the Compton scale turns out to be a miraculous scale. More recently,the author has shown that the background Zero Point fluctuations lead at the Compton scale to the electron's Quantum Mechanical behaviour including its spin and anomalous $g=2$ factor \cite{bg-ijtp->}.\\
Finally, we may point out that in the preceding section, we had tacitly assumed the value of the fine structure constant $\alpha.$ If on the other hand, we require the coefficient to be half to give the correct expression for the Compton length, then we can deduce the very nearly accurate value of $\alpha.$

\begin{flushleft}
\large {\bf ACKNOWLEDGEMENT}
\end{flushleft}
\noindent I would like to thank Prof.Walter Greiner of the Frankfurt Institute of Advanced Studies for valuable discussions.



\begin{thebibliography}{99}
\bibitem {bg-cu01} \textsc{Sidharth, B.G.} (2001). \emph{Chaotic Universe: From the Planck to the Hubble Scale}
(Nova Science, New York).
\bibitem {bgmg8-99} \textsc{Sidharth, B.G.} (1999). \emph{Proc. of the Eighth Marcell Grossmann Meeting on
General Relativity (1997)} Piran, T. (ed.) (World Scientific,
Singapore), pp.476--479.
\bibitem {bg-ijmpa98} \textsc{Sidharth, B.G.} (1998). \emph{Int.J. of Mod.Phys.A} 13, (15), pp.2599ff.
\bibitem {bg-ijtp98} \textsc{Sidharth, B.G.} (1998). \emph{International Journal of Theoretical
Physics} Vol.37, No.4, pp.1307--1312.
\bibitem {duffy} \textsc{Sidharth, B.G.}(2006). in \emph{Einstein and Poincare}
Dvaeglazov, V. et al. (eds.) (Apeiron Press, Canada).
\bibitem {nipr-89} \textsc{Nicolis, G. and Prigogine,} I. (1989). \emph{Exploring Complexity}
(W.H. Freeman, New York), p.10.

\bibitem {mwt} \textsc{Misner, C.W., Thorne, K.S. and Wheeler, J.A.} (1973). \emph{Gravitation}
(W.H. Freeman, San Francisco), pp.819ff.
\bibitem {uof} \textsc{Sidharth, B.G.} (2005). \emph{The Universe of Fluctuations} (Springer,
Netherlands).
\bibitem {ce-98} \textsc{Cercignani, C.} (1998). \emph{Found.Phys.Lett.} Vol.11, No.2, pp.189-199.
\bibitem {cegasc-72} \textsc{Cercignani, C., Galgani, L. and Scotti, A.} (1972). \emph{Phys.Lett.} 38A, pp.403.
\bibitem {huang} \textsc{Huang, K.} (1975). \emph{Statistical Mechanics} (Wiley Eastern, New Delhi), pp.75ff.
\bibitem {rohr}\textsc{ Rohrlich, F.,} \emph{Classical Charged Particles}, Addison-Wesley, Reading, Mass., 1965, pp.145ff.
\bibitem{bgs-tlsr->} \textsc{Sidharth,B.G.,}\emph{The limits of special relativity}, to appear in Foundations of Physics.
\bibitem {jica-99}\textsc{Jimenez, J.L. and Campos, I.} (1999). \emph{Found. of Phys.Lett.} \textbf{12}, 2, pp.127--146.
\bibitem {barut} \textsc{Barut, A.O.,} ``Electrodynamics and Classical Theory of Fields and Particles'', Dover Publications, Inc., New York, 1964, p.97ff.
\bibitem {bgs} \textsc{Sidharth, B.G.,} "The Lorentz Dirac and Dirac Equations" to appear in
the Journal, Electromagnetic Phenomena. Cf. also Sidharth, B.G.,
arXiv-physics/070237v1.
\bibitem {jc}\textsc{ Jimenez, J.L., and Campos, I.,} Found. of Phys.Lett., Vol.12, No.2, 1999, p.127-146.
\bibitem {diracpqm} Dirac, P.A.M.,  "The Principles of Quantum Mechanics", Clarendon Press, Oxford, 1958, pp.4ff, pp.253ff.
\bibitem {bgs-tu08} \textsc{B.G. Sidharth,} ``The Thermodynamic Universe '', World Scientific, Singapore, 2008.
\bibitem {hoyle} \textsc{Hoyle, F., and Narlikar,} ``Action at a Distance in Physics and Cosmology'', W.H. Freeman, New York, 1974, pp.12-18.
\bibitem {nar} \textsc{Hoyle, F., and Narlikar, J.V.,  }``Lectures on Cosmology and Action at a
Distance Electrodynamics'', World Scientific, Singapore, 1996,
pp.14ff.
\bibitem{bgs-fw->}\textsc{Sidharth,B.G.,} \emph{Feynman Wheeler perfect absorber theory in a new light} to appear in Foundations of Physics.
\bibitem{puthoff}\textsc{Puthoff,H.E.,} Int.J.Th.Phys., 2007, Vol.46, pp.3005-3008
\bibitem {motr-88} \textsc{Mostepanenko, V.M. and Trunov, N.N.} (1988). \emph{Sov.Phys.Usp.} 31,
(11), November 1988, pp.965--987.

\bibitem{dewi} \textsc{DeWitt,B.,In Sarlemijn,A., Spaarnay }(eds)(1989)\emph{Physics in the Making} (North Holland, Amsterdam )
\bibitem{veneziano}\textsc{Veneziano, G.} (1998). \emph{The Geometric Universe}
Huggett,S.A. et al. (eds.) (Oxford University Press, Oxford),
p.235ff.
\bibitem{bg-ijtp->}\textsc{Sidharth,B.G.,} \emph{Re-visiting Zitterbewegung} to appear in Int.Jour.Th. Ph.
\end{thebibliography}
\end{document}